# Adam Smith's Theory of Value: A Reappraisal of Classical Price Discovery


Sabiou M. Inoua and Vernon L. Smith

Chapman University



**Abstract**. The relevance of Adam Smith for understanding human morality and sociality is recognized in the growing interest in his work on moral sentiments among scholars of various academic backgrounds. But, paradoxically, Adam Smith's theory of economic value enjoys a less prominent stature today among economists, who, while they view him as the "father of modern economics", considered him more as having had the right intuitions about a market economy than as having developed the right concepts and the technical tools for studying it. Yet the neoclassical tradition, which replaced the classical school around 1870, failed to provide a satisfactory theory of market price formation. Adam Smith's sketch of market price formation (Ch. VII, Book I, *Wealth of Nations*), and more generally the classical view of competition as a collective higgling and bargaining process, as this paper argues, offers a helpful foundation on which to build a modern theory of market price formation, despite any shortcomings of the original classical formulation (notably its insistence on long-run, natural value). Also, with hindsight, the experimental market findings established the remarkable stability, efficiency, and robustness of the old view of competition, suggesting a rehabilitation of classical price discovery. This paper reappraises classical price theory as Adam Smith articulated it; we explicate key propositions from his price theory and derive them from a simple model, which is an elementary sketch of the authors' more general theory of competitive market price formation.




**1 Introduction**

The relevance of Adam Smith for understanding human morality and sociality is generally accepted, as registered in the growing interest that his work is stimulating among scholars of various academic backgrounds (philosophers, political theorists, sociologists, economists). But, paradoxically, Adam Smith's theory of economic value (1776 [1904], Book I) enjoys a less prominent stature today among economists, who, while they may view him as the "father of modern economics", consider him as having had the right intuitions about a market economy but not as having developed the right concepts nor the technical tools for studying it. The dominant historiography of economics draws indeed a picture of classical economics in which Adam Smith is overshadowed by the English classical followers (notably Ricardo), and it portrays the classical school itself as eclipsed in the 1870s by the neoclassical school. Yet the new school faces important and persistent difficulties. First, it failed to provide a satisfactory theory of market price formation owing to the dominant axiom of price-taking behavior; for if everyone takes prices as given, how do these prices emerge in the first place? Who is giving the prices? One early escape from this crucial price-discovery problem supposed that all traders should have complete information on supply and demand and the consequent equilibrium prices (Jevons, 1871 [1888]); the other, which stimulated general equilibrium theory, imagines a fictional auctioneer who finds the equilibrium prices by trial-and-error



adjustments or tatonnement (Walras, 1874 [1954]).[1] The difficulty of a neoclassical approach to price formation is compounded by a more technical aggregation problem established in the 1970s in an important theorem (by Sonnenschein, Mantel, and Debreu), which uncovers an intrinsic lacuna in the core principle of deriving economic regularities from individual utility-maximizing rationality; for the theorem shows that the demand of such agents is essentially arbitrary in the aggregate (Sonnenschein, 1972, 1973a, 1973b; Debreu, 1974; Mantel, 1974).[2] In contrast, economic regularities are better viewed as emergent properties of interacting agents in the marketplace as it is organically represented in the classical school. Finally, experimental economics established the stability, efficiency, and robustness of the market mechanism under conditions in which we should expect "market failures" according to the standard neoclassical theory: markets with a few traders, who know only their private valuations of the good, and who generate the prices through their bids and asks; yet, unaccountably, these markets converge to equilibrium and maximum efficiency (V. L. Smith, 1962, 1965; Plott, 1982; V. L. Smith, 1982; V. L. Smith & Williams, 1990).

Adam Smith's sketch of competitive price formation (Ch. VII, Book I, *Wealth of Nations*, 1776 [1904]), and more generally the old classical view of competition as a multilateral higgling and bargaining rivalry process, as this paper argues, offers a helpful foundation on which to build a modern theory of market price formation, despite any shortcomings of the original classical formulation (notably its well-known insistence on long-run, natural value). This paper is a reappraisal of classical price theory as Adam Smith articulated it; we explicate key propositions from his value theory and derive them from an elementary mathematical model (Section 7), which is a sketch of the authors' theory of competitive market price formation.

---

[1] "A market, then, is theoretically perfect only when all traders have perfect knowledge of the conditions of supply and demand, and the consequent ratio of exchange; and in such a market, as we shall now see, there can only be one ratio of exchange of one uniform commodity at any moment." (Jevons, 1871 [1888]). Walras merely sets the stage for the auctioneer story: historically, the explicit reference to the fictional auctioneer will be adopted later, to fill in the missing link in neoclassical price theory. For a thorough discussion on this matter, see the new translation of Walras by Walker and van Daal (Walras, 1874 [1896, 2014]).

[2] For a review of this important negative result of general-equilibrium theory, see Shafer and Sonnenschein (1982), Kirman (1989), and Rizvi (2006).

Four obstacles limit a modern appreciation of classical economics and recognition of Adam Smith's contribution to value theory. The first difficulty is the view that Adam Smith's formulation of price theory is superseded by some of his followers (especially Ricardo, whose reduced formulation is largely responsible for the dismissal of classical economics as a mere labor theory of value that ignored the demand side of price formation). The second difficulty is the "unsystematic" form of the classical discussions on value; for classical economics is largely rooted in astute observations of real economic phenomena, but organized in an informal, ostensibly unsystematic, way; but this does not make it less rigorous as emphasized throughout this paper. The third obstacle relates to the classical technical jargon (natural price, monopoly price, effectual demand, etc.), which to a large extent is outmoded today, and not always for good reasons (Section 6).

But by far the greatest difficulty relates to an equally outmoded tripartite articulation of value theory (Section 2), which is often confused in modern interpretations. Until Marshall, value theory tended to be concerned not only with market price formation (the main goal of value theory, about which there was in fact widespread consensus, in both the classical and neoclassical schools, that market price is regulated by the law of supply and demand, as emphasized in Section 2), but also with more philosophical, preliminary



investigations suggested by two other, more basic, problems of value theory—the measure of value (determining a universal and invariable standard of value) and the origin of value (the quest for the most primitive cause of value). The two other problems of value theory are essentially chicken-and-egg, metaphysical, issues that generated much of the controversies on value among the classical economists themselves, and the marginalists' opposition to the old school. It can be shown that the English classical economists' obsession with labor (and the later marginalists' insistence on marginal utility for that matter) is largely due to this scientifically peripheral issue, since labor was regarded as the closest to being the invariable and universal value standard and ultimate cause of value (to which view the marginalists' opposed marginal utility); yet none of the famous classical economists (except perhaps Ricardo) considered that this special status of labor implies that competitive market price formation amounts to a labor theory of value (no more than the early neoclassical authors' emphasis on marginal utility meant that they were disregarding cost and supply as regulator of market price). For Adam Smith the relevance of the labor theory of value was confined to a hypothetical early and rude state of society, namely a primitive barter economy, in which all labor skills are identical (hunting skills, for example), land is not appropriated, and capital is non-existent as a separate factor of production; thus, starting from chapter VII, Adam Smith expounded the theory of price formation from buyer-buyer and seller-seller competition expressed in their aggregated supply and demand, which is relevant for a modern economy (Section 5).[3] In fact, Ricardo is an exception in the classical school in his attempt to generalize the labor theory beyond the primitive case. Both Say and Malthus restated classical price theory in the spirit of Adam Smith's formulation. But Ricardo had more disciples than the other classical economists!

---

[3] The labor theory of value is equivalent to assuming a Leontief price system, as is known since the influential revival of Ricardo's theory by Sraffa (1960). Section 7 (Theorem 2) contains a slightly different, simple, derivation of the labor theory of value in which the profit rate does not appear explicitly, since profit is already included in cost of production, as it should be strictly speaking in the original classical treatment.

A systematic historical account of this complex articulation of value theory is not possible here, for it would amount in effect to a history of value theory itself; thus, we document in the next section a few major aspects of the three problems of value theory and discuss in greater detail in Section 5 how Adam Smith dealt with them.

## 2 The Three Problems of Value Theory

Adam Smith announced his articulation of value theory as follows:

> "In order to investigate the principles which regulate the exchangeable value of commodities, I shall endeavour to shew, First, what is the real measure of this exchangeable value […]. Secondly, what are the different parts of which this real price is composed or made up. And, lastly, what are the different circumstances which sometimes raise some or all of these different parts of price above, and sometimes sink them below their natural or ordinary rate; or, what are the causes which sometimes hinder the market price, that is, the actual price of commodities, from coinciding exactly with what may be called their natural price. I shall endeavour to explain, as fully and distinctly as I can, those three subjects in the three following chapters, for which I must very earnestly entreat both the patience and attention of the reader […]."

Following this plan of investigation more or less literally, value theorists until Marshall tended to be concerned with three problems: (1) the measure of value (determining a universal and invariable standard of value); (2) the origin of value (between utility and cost, which is the ultimate, most primitive cause of value?); (3) price theory proper, or the problem of market price formation, to which Marshall reduced value theory. The three problems (or four, if we count the preliminary discussion on the nature of value) can be easily identified in the treatises on value from Smith to Marshall. For example, Malthus's chapter on value in *Principles of Political Economy* (1820 [1836]) is "On the



Nature, Causes, and Measures of Value" (Bk. I, Ch. II, p. 50); Section I is "On the different sorts of Value" (notably value in use and value in exchange); Section II is "Of Demand and Supply as they affect Exchangeable Value", or the theory of competitive market price formation; Section III, "Of the Cost of Production as affected by the Demand and Supply, and on the Mode of representing Demand" derives from the general theory of supply and demand the doctrine according to which price tends to correspond to cost in the long run; finally Sections IV-VII are a lengthy discussion dealing with the difficult problem of an invariable value standard.

The first two problems (the quest for an ultimate cause of value and an absolute measure of value) are, as we said earlier, fundamentally chicken-and-egg problems, not to be confused with price theory proper. Alfred Marshall did not of course invent the supply-and-demand theory of price formation, not even its famous diagram (Ekelund Jr & Thornton, 1991): his pair of scissors' metaphor was merely intended to emphasize the futility, from the viewpoint of price formation theory, of the "doctrines as to the ultimate tendencies, the causes of causes, the *causae causantes*" of value (which in the classical school was investigated in terms of the relations between cost of production and value), because, argued Marshall, both utility and cost are mutually causing or determining market price (1890 [1920], pp. 348, 821). That is, the price mechanism is a complex of mutually reinforcing causes, rather than the unidirectional chain of causation put forward by the various doctrines on the origin of value. Marshall illustrates this point on Ricardo's argument that labor is the ultimate cause of value (which overlooks in market price formation, the fact that "the various elements govern one another mutually, and not successively in a long chain of causation", p. 816) and on Jevons' argument for



marginal utility (that substitutes "a catena of causes for mutual causation", p. 818, marginal summary).

That both utility and cost determine together market price is an obvious fact that no economist before Marshall denied. Utility and cost are the two basic causes of value. The problem of the origin of value is a debate over the original or most primitive cause of value. For Adam Smith, labor was this origin of value: "Labour was the first price, the original purchase-money that was paid for all things." (1776 [1904], bk. I, ch. V, p. 32) In the classical school, the controversy over the origin of value will oppose Ricardo, who followed Adam Smith's view on the "original source" of value (1817 [1821] p. 5), and J.-B. Say, who held utility as the "foundation of value" (e.g. 1828 [1836], part I, div. I, ch. III). J.-B. Say granted that a good should involve some toil and trouble to make, if it is to command any value (for otherwise none will pay a penny for it); but he pointed out that none would have accepted to suffer this labor unless the object was useful in the first place: so utility causes labor, so to speak, in which case it is more fundamental; thus he regarded utility as the foundation of value. This does not conclude the quest for the ultimate cause of value, however, which could go on endlessly; for underlying Ricardo's theory, or even contributing to it, is another way of seeing the problem of the origin of value, which goes back to Adam Smith and which is equally difficult to refute: a regression to the ultimate, primitive cause of value going from the present to the remotest past (Ricardo, 1817 [1821] p. 18). Here is more specifically how this metaphysics goes. A good is produced by labor, which is assisted by implements (or capital), which are themselves made by labor and other implements, which are themselves made by labor and other implements.... One could postulate that this chain of causes ultimately leads to labor as "the first price" that was "paid for all things", as Adam Smith speculated, or



"the original source" of value, as Ricardo insisted. But such a problem is beyond the scope of positive economics.

In summary, the controversy on the origin of value consisted in identifying, not the exclusive cause of value (utility or cost), but the original or most ultimate of the two causes: labor (the English classical tradition, notably Adam Smith and Ricardo), utility (the French classical tradition, notably J.-B. Say), or marginal utility (the marginalist movement in the 1870s).

Consider, for example, the Austrian marginalists, who also adopted the old approach to value theory in terms of the three main problems, typically articulated in two parts, however (Menger, 1871 [1950], ch. III versus ch. V; Böhm-Bawerk, 1888 [1891], bk. III versus bk. IV). The first part ("value theory" narrowly defined) deals with the preliminary, more philosophical problems about the nature and origin of value, and this is wherein lies in the Austrian literature the well-known, relatively more radical, insistence on the subjectivity of value and the centrality of marginal utility, in reaction to the English classical school's argument in favor of labor or cost; yet no Austrian marginalist denied the importance of cost and supply in market price formation (or "price theory"), which is the second main part of the Austrian articulation of value theory, and for which the Austrian view from the beginning adopted what we shall emphasize to be the old, if implicit, view of supply and demand, namely that centered on traders' monetary valuations of a commodity (as opposed to the abstract utility representation in terms of pleasure that Jevons and Walras inaugurated, or the even more abstract preference-relation view of the modern formulation). On the important problem of price formation, in other



words, the Austrian tradition built on the old, classical, view on competition.[4] Value is subjective in the Austrian literature, not in the sense that the consumer marginal utility (or valuation) is the sole operational cause of value, but because the other proximate cause, production cost (or the seller's willingness to accept a price) is itself viewed in this tradition as an ultimately subjective reality, decided in final analysis by the consumer's valuation of the product to be sold:

> "And now we have to consider the causal connection which has ended in this price. It runs, in the clearest possible way, in an unbroken chain from value and price of products to value and price of costs--from iron wares to raw iron, and not conversely. The links in the chain are these. The valuation which consumers subjectively put upon iron products forms the first link. This helps, next, to determine the figures of the valuation--the money price at which consumers can take part in the demand for iron products. These prices, then, determine, in methods with which we are now familiar, the resultant price of iron products in the market for such products. This resultant price, again, indicates to the producers the (exchange) valuation which they in turn may attach to the productive material iron, and thus the figure at which they may enter the market as buyers of iron. From their figures, finally, results the market price of iron." (Böhm-Bawerk, 1888 [1891], p. 226)

Walras and Jevons proposed to the English classical school a similar argument in favor of marginal utility as the origin of value: "Cost of production determines supply. Supply determines final degree of utility [namely, marginal utility]. Final degree of utility determines value." (Jevons, 1871 [1888], p. 165)[5] Put differently, marginal utility is not only the origin of pure-exchange value (Walras, 1874 [1896, 2014], Lesson 16), but it remains the ultimate cause of value even when production is considered (Walras, 1874 [1896,

---

[4] In this respect, the most significant shift that occurred during the marginal revolution was when Jevons and Walras, following Cournot's innovation (1838 [1897], ch. VIII), redefine competition as passive price taking-behavior, a reformulation that fits well the conception of supply and demand as optimal quantity choices at given prices, but which is hardly compatible with rivalrous "higgling and bargaining" market behavior.

[5] In fact, labor is no conceptually distinct substance in Jevons's value theory, since labor being cause of pain, it is just a modality of utility, a negative utility, or disutility: utility being identified with pleasure, there is "equivalence of labour and utility" (Jevons, 1871 [1888], p. 177)



2014], Lessons 17-18), for the determination of production cost (equilibrium in the market for labor, capital, and land services) depends on the price of the produce, itself determined in final analysis by the consumer's marginal utility.

The value standard problem also led to metaphysical controversies that also tended to overshadow the theory of competitive market price formation. As said earlier, the English classical economists (Smith, Ricardo, Malthus) concurred on the special status of labor not only as being the ultimate cause of value but also as being the closest to being the invariable and universal value standard; yet an obscure controversy will oppose Malthus to Ricardo as to the specific labor form that should serve the role of an invariable value standard (the famous labor commanded versus embodied controversy). Ricardo, as it is well known, suggested that the measure of value should be the labor it cost to produce a good (or labor embodied) whereas Malthus proposed instead the labor that the good commands in a market (or labor commanded). (Adam Smith evoked both measures in his speculation on the rude state of society, but treated them equivalently, as they are by construction in this hypothetical society.) The controversy between Ricardo and Malthus remained unsettled; Ricardo, in particular, who considered this problem to be "the most difficult question in Political Economy", wrestled with it till his last days (1817 [2004], Sraffa's introduction, p. xl). [6] The difficulty is intrinsic indeed and the controversy may well be as a later commentator portrayed it, "the chimera of an invariable standard of value".[7] For it is not difficult to see with hindsight that the problem of

---

[6] Letter to Malthus, 3 August 1823 (*Works*, vol. IX, p. 325); cited in *David Ricardo: Notes on Malthus's 'Measure of Value'*, Introduction, p. xvi.

[7] E. Cannan, *A Review of Economic Theory* ([1929] 1964)*, cited in Ricardo (1817 [2004], Sraffa's Introduction, p. xl).



value standard is hardly solvable under Ricardo's premise that the invariable measure, rather than being a precondition to value theory, ought to be decided based on a theory of value: "Is it not clear then that as soon as we are in possession of the knowledge of the circumstances which determine the value of commodities, we are enabled to say what is necessary to give us an invariable measure of value?" (p. xli) This is indeed a chicken-and-egg problem. A theory of value that presupposes a value standard cannot determine the value of this standard, which is arbitrary therein (and considered as unity only for convenience); but then the value of any other good, or combination of goods, being given in terms of this standard, cannot be absolutely invariant in this theory (since it can only be invariable relative to the standard). On the other hand, if there can be a valid theory of value that requires no standard of value beforehand, then there was in fact no problem of standard in the first place. A value standard is of course no more than a convenient convention; and money, adjusted for inflation, seems to be the natural choice, and the one Adam Smith falls back on, after a complex investigation briefly emphasized below (Section 5). J.-B. Say avoided altogether the quest for an invariable measure of value, which he considered to be the economics' equivalent of the old problem of the squaring of the circle ("la quadrature du cercle de l'économie politique") and the quest of which he regarded as sterile (1828 [1836], part I, div. I, ch. II, p. 39). J.S. Mill reached the same conclusion: "There has been much discussion among political economists respecting a Measure of Value. An importance has been attached to the subject, greater than it deserved, and what has been written respecting it has contributed not a little to the reproach of logomachy, which is brought, with much exaggeration, but not



altogether without ground, against the speculations of political economists. It is necessary however to touch upon the subject, if only to show how little there is to be said on it." (1848 [1965], bk. III, ch. XV, §1, p.577)

Thus was, briefly speaking, the old articulation of value theory in three problems. The important point to keep in mind in the sequel is that which we tried to emphasize throughout the ongoing discussion: While the development of value theory tended to be obscured by a great deal of metaphysical controversies related to the measure and cause of value, there was widespread consensus across both schools on the fact that market price is determined in a market competition of supply and demand (hence by both utility and cost): "when prices are said to be determined by demand and supply, it is not meant that they are determined either by the demand alone, or by the supply alone, but by their relation to each other." (Malthus, 1820 [1836], p. 62). "Almost all writers have agreed substantially, and have rightly agreed, in founding exchangeable value upon two elements: power in the article valued to meet some natural desire or some casual purpose of man [utility], in the first place, and, in the second place, upon difficulty of attainment [cost]. These two elements must meet, must come into combination, before any value in exchange can be established." (De Quincey, 1844, p. 13) The popular simplification of the history of value theory in terms of a supply-side, cost or labor value theory (classical school), superseded by a demand-side marginal-utility value theory, and superseded in turn by Marshall's synthesis of the two views, is therefore a misreading. Marshall's synthesis is more subtle.

## 3 Marshall's View on Adam Smith and the Nature of Marshall's Synthesis

Of all the commentators on the intellectual history of economics, Alfred Marshall is perhaps the author who most clearly understood the pivotal contributions of Adam Smith to modern economics. He regarded Adam Smith as having launched an epoch in economics when he built, from a core methodological principle overlooked in modern commentaries, a value theory that unifies all of economics ([1890] 1920, Appendix B, p. 627). This principle consists of dealing, as regards individual economic decisions, not directly with the unobservable ultimate psychological forces driving them (need, desire, pleasure) but with the *monetary sacrifices* that people make to satisfy them: formally, their reservation prices. Thus, the relevant concepts for demand and supply theory are the (maximum) money prices consumers are willing to pay and the (minimum) money prices suppliers are willing to accept in the marketplace. This is a most fundamental classical principle that Marshall incorporated into his reformulation of neoclassical value theory in contrast to the hedonistic marginal utilitarianism of Jevons and Walras, who make pleasure the fundamental motivating category of economics. Alfred Marshall, perceptively recognizing this classical methodology, credited its discovery to Adam Smith, distinguished from predecessors and successors "by a clearer insight into the balancing and weighing, by means of money, of the desire for the possession of a thing on the one hand, and on the other of all the various efforts and self-denials which directly and indirectly contribute towards making it. Important as had been the steps that others had taken in this direction, the advance made by him was so great that he really opened out this new point of view, and by so doing made an epoch." (1890 [1920], Appendix B, p. 627). It is in fact this principle for measuring motives that confers upon economics a



special quantitative nature among the social sciences (1890 [1920], Book I, Ch. II, p. 12). The willingness-to-pay/willingness-to-accept approach to supply and demand frames value theory throughout the classical literature (Inoua & Smith, 2020a, 2020b); it is also adopted, not only by Marshall, but also by the Austrian marginalists in their explanation of competitive market price formation. In the 1950s, moreover, experimental economists, inspired by Marshall's treatments of this classical principle, adopted it in their implementation of supply and demand functions (Chamberlin, 1948; V. L. Smith, 1962).

Marshall's synthesis of classical and neoclassical value theory is therefore methodological. Though he accepted diminishing marginal utility as central to value theory (making him a marginalist of course), yet he saw in Jevons's program a major setback from the core methodological principle of classical economics. Marshall's synthesis, in other words, is more subtle than suggested by the popular interpretation of his "pairs of scissors" metaphor, which, as noted earlier, Marshall merely intended to emphasize the futility, from the viewpoint of price theory, of the old quest for the ultimate cause of value. [8]

## 4 Adam Smith Belittled as an Economist

Unlike Marshall, however, influential commentators on the history of economics tended to diminish, even belittle, Adam Smith's technical contributions to value theory. A brief review is enough to show the extent to which certain influential historians of economics

---

[8] Marshall put an end to the quest for the absolute of value only in effect, however, by reducing value theory to its positive part (namely price theory) rather than by solving the metaphysical controversies. For these controversies have no conclusion. In fact, be it emphasized in passing, Marshall's classical re-foundation of price theory, in terms of the monetary sacrifices people make to meet their wants, can itself be invoked in favor of the English classical view that labor is the ultimate cause of value, only with labor generalized to effort (whether physical effort or monetary sacrifices).



have denigrated the classical school more generally as they interpret it in neoclassical terms.

J. Schumpeter opined that: "There is no theory of monopoly [in *Wealth of Nations*]. The proposition […] that 'the price of monopoly is upon every occasion the highest which can be got' might be the product of a not very intelligent layman—taken literally, it is not even true. But neither is the mechanism of competition made the subject of more searching analysis. In consequence, A. Smith fails to prove satisfactorily his proposition that the competitive price is 'the lowest which the sellers can commonly afford to take'—to the modern reader it is a source of wonder what kind of argument he took for proof. Still less did he attempt to prove that competition tends to minimize costs, though it is evident that he must have believed it." (Schumpeter, 1954 [2006], p. 294) Likewise, G. Stigler, in his historical essays, portrayed the classical concepts of utility and competition as archaic versions of their modern formulations (Stigler, 1957, 1982). He viewed Adam Smith, not as the author of a unified theory of value, but as "a manufacturer of traditions", one of which is simply "to pay no attention to the formal theory of monopoly" (Stigler, 1982). M. Blaug went further and concluded that "Adam Smith had no consistent theory of wages and rents and no theory of profit or pure interest at all. To say that the normal price of an article is the price that just covers money costs is to explain prices by prices. In this sense, Adam Smith had no theory of value whatever." (Blaug, 1985, p. 39) Thus, it has become a common critique of Adam Smith that he held at best a confused view on value and income distribution. This misreading, we believe, is not solely induced by a neoclassical interpretation: it is reinforced by a Ricardian reading of Adam Smith, which represents a lighter but similar bias: That is, the premise (or preju-



dice) that Adam Smith had a theory of income distribution à la Ricardo, one that is separate from value theory and primary with respect to it. Hence many commentators failed to see that Smith's views on wages, profits, and rents are consequences of his price theory, as it is articulated in Ch. VII.

## 5 The Articulation of Smith's Value Theory

We return now to the original inspiration of much of the development on value discussed previously—Adam Smith's conceptualization of value in his magnum opus, Book I (whose first part, Ch. I-III on the 'division of labor', pertains to economic development).

Chapter VII is the most incisive on value theory, for it presents the general theory of market price formation, which Smith then applies, in Ch. VIII-XI, to explain the wages of labor (Ch. VIII and X), the profits of capital (Ch. IX and X), and the rents of land (Ch. XI). The previous chapters on value, Ch. IV-VI, primarily serve as preparatory discussions on the nature of value, which starts with the fundamental distinction between "value in use", or the value a person attaches to a good in view of the good's utility, and "value in exchange", or the ratio at which a good exchanges for another (p. 30). (A convention throughout the classical school consists of using the term "value", without qualification, in reference to exchange-value.) Adam Smith then tackled the tricky problem of the standard of value (Ch. V). When value is given in terms of this common measure, it is classically known as "price", which is a much simpler notion (and so familiar that the technical nature of its logical origin is easily forgotten); then value theory becomes price theory. The problem then is to identify a medium that can serve as a value standard. Adam Smith first framed this problem in the most abstract way. He wanted a medium



for comparing "the values of different commodities at all times and at all places" (p. 38). Of course, this standard should be itself stable in value if it is to indicate the "real price" of commodities. Adam Smith had already investigated the "origin of money" (Ch. IV), for money being the universal medium of exchange and unit of account in modern economies, it is the natural value standard; but money being variable in value, "money price" or "nominal price" is a poor indicator of "real price" over long periods. Moreover, money cannot serve as the standard for "all times", for it was not used in the primitive state of society: barter, according to Adam Smith, was the primordial type of exchange (Ch. IV).

There is in such speculation a clear temptation towards metaphysics, which Adam Smith did not always resist, and which, as we saw in Section 2, obscured much of the later development on value theory for modern readers. For by "all times" he literally included the "early and rude state of society"; under this absolute requirement, it is easy to see that no medium except labor can be used as standard, for labor is the only resource that is common to all exchangeable goods at all places and all times, including the hypothetical moneyless era, as is clear in the passage quoted earlier: "Labour was the first price, the original purchase-money that was paid for all things. It was not by gold or by silver, but by labour, that all the wealth of the world was originally purchased." (p. 32) Thus the problem of the standard, or measure, of value led Adam Smith to consider *the origin of value*, which would become a second important topic in value theory. Continuing, we can go further and consider labor as the absolute origin of exchange-value; for whatever the primordial exchange was in human history, it must have indeed involved labor; assume, for example, that the first economic act in human history was picking a fruit: this then was the first time that the phenomenon of exchange-value emerges—someone has endured some labor in exchange for a fruit. The problem of the value standard, in



other words, pushed Adam Smith into metaphysical beginnings: the origin of money, value, society, or even humanity.[9] This is not to say that such discussion is uninteresting, irrelevant, or false; but merely to emphasize that it is an order of enquiry beyond the *science of value*, for its falsifiability cannot be subjected to a positive test; hence we counted this problem as part of the *metaphysics of value*, which as we saw regroups the most difficult problems on value, which triggered major controversies that involved most economists from Smith to Marshall, and whose second aspect is the controversy over the ultimate cause of value. For the modern reader, the difficulty in reading these controversies is compounded, because the different aspects of value are not often clearly distinguished.

Fortunately, there is a clear enough demarcation between these two orders of investigation in Adam Smith's book. The metaphysics of value is mostly concentrated in Chapters IV-VI, whereas the science of value truly begins in Ch. VII (and any note thereafter on the original conditions of humankind is passing and merely said by way of progression from the simple to the complex). Adam Smith's pragmatism, moreover, eventually outweighed his metaphysics even in his preliminary speculations. The quantity of labor, he noted, is an abstract resource which is not operational in ordinary transactions and is not easy to measure because of its heterogeneity. In practice, money price (or monetary valuation) regulates almost the totality of ordinary economic life. Yet Adam Smith needed also, as we would say today, a way of controlling for inflation when the very long run is considered. Hence, his long digression (in Ch. XI) on the variations of the currency—primarily silver—whose real value he assessed in terms of corn, whose price, which was then available for four centuries back, Adam Smith believed to be stable from century to century, though it fluctuates from year to year. Thus, he measured the real price of silver by the amount of corn that this currency can buy. Smith was sensitive to measurement issues handled today by the technique of index numbers.[10]

---

[9] Methodologically, "Smith believed that writers of 'didactical' discourse ought ideally to deliver a system of science by laying down 'certain principles, known or proved, in the beginning, from whence we account for the several phenomena, connecting all together by the same chain'... Smith drew an implicit distinction between the method used in expounding a system of thought and that employed in establishing such a system...In short, the task of establishing a system of thought must be conducted in terms of the combination of reason and experience..." (Smith 1795 [1980], p. 1) This commitment to principles of reason based on experience marks his style, for in Smith (1759 [1853]) is articulated a theory of the origins and function of human sociability that give rise to the two pillars of society: Beneficence, which underlies reciprocity and social exchange; and Justice as security from injury and thus essential to property. Its relevance for all time, is articulated and applied to contemporary behavioral experiments by Smith and Wilson (2019).

Smith ended his preliminary discussions with a simple accounting of value, or "the component parts of price" (Ch. VI). The sustainability of the price of a good requires it to cover the sum of the wages, profits, and rents that reward the three agents that produce this good—land, labor, and capital. So once price is explained in general, and wage and rent by implication, profit follows residually; thus, if Smith explained the profit rate in general from the competition of capital (or "stock"), he at times derived the overall pattern of profits directly from that of wages, which tend to evolve inversely (Ch. IX). Throughout the classical literature, the idea that price corresponds to cost is considered in two senses, depending on the context, and this ambivalence is misleading if not kept in mind. The first one is the obvious accounting identity just noted: rent is the cost associated to land; wage is that associated to labor, and profit (which, as emphasized below, is classically viewed as a cost) is that associated to capital. Thus "price equals cost" here is a mere truism.[11] The second meaning corresponds to a theoretical proposition of utmost importance in this school: it says that *price converges to minimum cost under free competitive entry*. As generalized and applied to the whole economy, this proposition holds when the competition of landowners, of workers, and of capitalists is so intense that the rent, wage, and profit rates are the minimum they can be, and hence prices correspond to minimum costs.[12] But the proposition also applies locally, to a given market, when the competition on the side of supply is intense and only the most efficient suppliers succeed to sell the good at the lowest possible price. We return to this key proposition below.

---

[10] On this issue, see also the analysis in Hoover and Dowell (2001). The authors thank an anonymous referee for pointing out this reference.

The "market price" of a good is regulated by competition of supply and demand. When, moreover, the competition on the side of suppliers is free, in the sense of being unconstrained, and the most intense, the market price converges to the "natural price", the lowest price at which the good can sell, and continue to be produced and brought to market. Classically, "cost" includes a "normal" or "ordinary" profit expectation. So technically, cost is classically a synonym for (long-run) *supplier reservation price*, and by price convergence to cost or natural price, the classical economists meant, not convergence to zero profit, of course, but to zero *surplus* above the overall minimum acceptable profit, or the natural profit rate. Smith, and all his disciples, insisted on *free competition* because they viewed it as a norm, and in two senses: it is classically the ideal case, the socially optimal state under which prices are so low that consumers of all orders of society can afford it (in short, a state of cheapness and plenty, in particular for basic needs). But the classical economists also considered it to prevail reasonably in practice, and to be expected, if market supply is not artificially restricted. Even under Mercantilist interventions, physical, and other barriers, it is the natural history and course of a market *in the long run*. All limitations to competition, whether natural or artificial, are collectively referred to as "monopoly", a term whose classical meaning does not map into modern terminology. Yet this classical understanding of monopoly was a standard one throughout scholastic economics (De Roover, 1951). The scholastic influences on Adam Smith's theory of value is a subject of utmost importance for a deep understanding of classical economics more generally, although space forbids to elaborate here on this fascinating topic, which is now fairly accessible thanks notably to Raymond de Roover's brilliant rehabilitation of this noble tradition of the scholastic doctors (De Roover, 1951, 1955, 1958, 1971).

---

[11] It is a mistake therefore to read this mere accounting decomposition of price as a price theory (a so-called 'adding-up value theory') and treating Adam Smith's text as a confusion of switching among various value theories from paragraph to paragraph.

[12] In a modern treatment, of course, we would say that the risk-adjusted return to investment in land, human capital and capital facilities must be equal.

The early neoclassical economists, following a distracting innovation by Cournot, reduced monopoly to its etymology: a market supplied by a lone seller. But for classical economics a lone seller may be the most efficient supplier of the good, who undersells all rivals and thus brings the price to its natural level; the etymological identification of monopoly deflects from the classical conception of markets as a process of adapting to



the conditions of supply as well as demand.[13] Under monopoly, as classically understood, namely under a limitation of competition among suppliers—as in the case of only one diamond mine—the market price reflects the condition that only the buyers willing to pay the most make purchases, and in this sense the price is the "highest to be got." For other goods, such as iron and water, more sellers enter, price is lower and lower-value buyers enter the market. This rich variety of outcomes were driven by natural market processes.

This theory is simple but rich in its implications. First, it is general, in that it applies to any competitive market, whatever the number of buyers and sellers that are involved in it (the smallest market being an isolated buyer-seller haggling): a most mischievous distortion of the classical view on competitive price formation is the neoclassical multitude of price theories based on the number of sellers in a market (monopoly, duopoly, oligopoly,…, "perfect competition"). In contrast, as sketched by Adam Smith, classical competition is a general process that operates on both sides of the market (buyer-buyer and seller-seller rivalry):

> "When the quantity of any commodity which is brought to market falls short of the
>
> effectual demand, all those who are willing to pay the whole value of the rent, wages

---

[13] "The economic theory of monopoly has been furnished in mathematical form, which is the clearest and most precise form, by Cournot ([1838] 1897, Chapter V) and by Dupuit […]. Unfortunately, economists have not thought it pertinent to learn about that theory, and have been reduced, on the subject of monopoly, to a confusion of ideas that, in their work, is perfectly expressed by a confusion of terminology. They have given the name of monopoly to economic activities that are found to be, not in the hands of one firm, but in the hands of a limited number of them. They have even given, by analogy, the name of monopoly to the ownership of certain productive services that are limited in quantity; for example, to the ownership of land." (Walras, 1874 [1896, 2014], pp. 442-443). It is to emphasize the severity and unfairness of this critique of the old view of competition and monopoly that Bertrand (1883, p. 503) emphasized the inconsistencies in Cournot's innovation, in which the very concept of competition as outbidding and underselling process is lost. As is clear from the context, Bertrand was not opposing to Cournot a new theory of competition (recentered on price as decision variable) but merely was emphasizing the superiority of the old view of competition.



and profit, which must be paid in order to bring it thither, cannot be supplied with the quantity which they want. Rather than want it altogether, some of them will be willing to give more. A competition will immediately begin among them, and the market price will rise more or less above the natural price, according as either the greatness of the deficiency, or the…eagerness of the competition." (A. Smith, 1776; 1904, Vol 1, p. 58)

"When the quantity brought to market exceeds the effectual demand, it cannot be all sold to those who are willing to pay the whole value of the rent, wages and profit, which must be paid in order to bring it thither. Some part must be sold to those who are willing to pay less, and the low price which they give for it must reduce the price of the whole. The market price will sink more or less below the natural price, according as the greatness of the excess increases more or less the competition of the sellers, or according as it happens to be more or less important to them to get immediately rid of the commodity."[14] (A. Smith, 1776; 1904, Vol. 1, p. 59)

This old view of competition developed by Adam Smith and clarified by the other classical economists is the natural foundation on which to build a theory of competitive market price formation, provided we generalize it to include buyer-seller competition.

Smith's exposition is rigorous throughout, except for one conceptual lapse that we address. At times, he treated the natural price as if it were synonymous with equilibrium price, or even the price attractor in all markets; hence his reference to the natural price at times as the "normal price" or "ordinary price", as if free competitive entry were a norm in the strong sense of being the normal state of affairs in all cases; hence also his oft-quoted yet misleading gravitation metaphor, namely that market price gravitates around the natural value; and finally his restricting demand to "those who are willing to pay the natural price", in his explanation of how the price returns to equilibrium in response to an excess demand at this natural price. This would be benign if he were throughout assuming free competitive entry, or long-run value exclusively, which is obviously not the case (and evident in his oranges versus scrap iron example).

---

[14] A. Smith, ever testing his model against observation, distinguishes perishables from inventories of durables, going on to add: "The same excess in the importation of perishable, will occasion a much greater competition than in that of durable commodities; in the importation of oranges, for example, than in that of old iron." (A. Smith, 1776 [1904], Vol. 1, p. 59) A. Smith also makes clear that his thinking about market processes is not confined only to long run supply and the "natural price."

Let it be noted that the classical solution to the paradox of value, which was long known before Adam Smith to be solved by the concept of scarcity (see formal definition below, Section 6), was explained passingly by Smith to his students in his *Lectures on Jurisprudence* (1763 [1869], p. 177). Water belongs to the class of goods which, by their abundance, usually involve little or no competition to acquire; so, though it is a vital good, its market price is close to its "natural value" (its minimum price), which is relatively low. A diamond, in contrast, has a much higher price by its rarity: its possession involves intense competition at the top of the distribution of buyers' valuations (or willingness to pay), so its price is near its "monopoly value" (the maximum price), as would be achieved in an auction for example. Adam Smith did not judge it worthwhile to elaborate on the famous water-diamond paradox in his magnum opus, because its solution was already commonplace: he was merely illustrating through this paradox the two meanings of value: value in use versus value in exchange.

**6 The Technical Jargon of Chapter VII**



Smith's Chapter VII introduces several technical terms, some alluded to above (cost, natural price, effectual demand), that he and others have used in different senses and meanings. We discuss them more fully here since this polysemy has created some confusion or controversy among the classical economists themselves. Largely due to Adam Smith, this polysemy limits an incisive penetration of classical price theory. For the sake of modern theory, it is perhaps best that one adheres closely to classical concepts or their formal equivalents, while avoiding some of their outmoded jargon. We suggest the following glossary, roughly translating from classical terms into modern versions.

| Classical term | Modern version |
| --- | --- |
| Use-value | Consumer valuation (formally, buyer's reservation price) |
| Exchange-value | Price |
| Cost | Seller's reservation price |
| Natural price | Minimum price |
| Monopoly price | Maximum price |
| Absolute demand (decided by need or desire) | Quantity needed |
| Effectual demand (decided by need, constrained by wealth) | Quantity demanded |
| Free competition | Maximum competition (entry freedom) on the supply side |

Here is a sketch of the different meanings attached to these and other related terms, in alphabetical order.

*Cost*. Classical cost includes profit: Cost = Prime Cost + Profit = Wages + Rent + Profit. Thus, the proposition Price = Cost is strictly speaking an accounting identity: Price = Wage + Rent + Profit. But Price = Cost has a second theoretical meaning and a central role in classical value theory in which the competition of firms drives the profit to the minimum necessary—the natural profit rate—for sustaining the supply of goods at a



price that satisfies consumer demand: price converges to minimum cost under free competition.

*Effectual demand.* "It is different from the absolute demand. A very poor man may be said in some sense to have a demand for a coach and six; he might like to have it; but his demand is not an effectual demand, as the commodity can never be brought to market in order to satisfy it." (A. Smith 1776 [1904], Book I, Ch. VII, p. 58) Absolute demand corresponds to the quantity needed or desired, independently of ability to pay (it is demand if price is hypothetically zero); *effectual demand*, in contrast, corresponds to quantity (effectually) demanded, which is constrained by wealth. Thus, *effectual demand* is none other than the classical equivalent of the modern notion of *demand*, considered as wealth (not income) constrained. The classical qualification is no longer needed since absolute demand falls in disuse. So far, so clear. But, unfortunately, Adam Smith first uses effectual demand, not in the general sense he intended, but in a specific context explained below (see natural price): thus *"effective demand"*, in the first occurrence, is restricted to "the demand of those who are willing to pay the natural price (see definition)"; then Adam Smith goes on to give the general idea he intended. J.S. Mill clarifies the matter as follows:

> "But what is meant by the demand? Not the mere desire for the commodity. A beggar may desire a diamond; but his desire, however great, will have no influence on the price. Writers have therefore given a more limited sense to demand, and have defined it, the wish to possess, combined with the power of purchasing. To distinguish demand in this technical sense, from the demand which is synonymous



with desire, they call the former *effectual* demand." (Mill, 1848 [1965], Book III,

Ch. II, §3, p. 465)

*Free competition.* The market price of a good varies between two limits, the minimum

willingness to accept across sellers and the maximum willingness to pay across buyers,

depending on the degree and type of competition in it. When competition is broadly the

most intense on the supply side, then the lowest-cost firms and the most efficient pro-

ducers undersell all rival firms, and the good sells at the lowest price, which is called the

"natural price": this case is classically known as free competition, because when no con-

straint (natural or artificial) limits the entry of firms and therefore their aggregate ca-

pacity to supply the good, competitive entry drives the price to its minimum. On the

other extreme is maximum competition on the demand side, in situations of extreme

scarcity, whereby the highest-value buyers outbid all the rival buyers to gain the few

units available to sell; then the price converges to the maximum willingness to pay, or

"monopoly price". Formally, the supply strike price is near the demand strike price. The

classical concept of *monopoly* is a very profound one, which Walras and other marginal-

ists (following Cournot's unfortunate innovation) misread, dismissed, and replaced with

the literal meaning of the term—the situation of a market supplied by a lone seller, in-

sulated from competition—which fails to capture the conceptual case of a single seller

as the most efficient firm that defeats all rivals by underselling them. Hence it is the

result of maximum competition on the supply side. Classically, monopoly simply means

the state of a market whose supply is so scarce that the price of the good is the maxi-

mum it can be. Where the extreme scarcity is natural (such as a diamond or a picture by

an old master), we have a properly defined natural monopoly. Classical competition is a

process not an outcome. Hence the observed outcome that a mature industry has only



one or two firms may mean that the competition and efficient size of firms is so intense that only one or two firms survived; or that, there are naturally only one or two sources of supply as with diamonds, or, finally, it may simply reflect state exclusionary grants which authorize only one or two firms.

The classical economists saw the greatest evil, and were critical of the last listed case, residing in artificial monopoly, wherein the scarcity is artificially created through some restriction of supply or entry, whereby a seller or a group of colluding sellers (often protected by state-granted mercantilist privileges) restrict the supply of a good and entry into the market. Hence, conditions that raise the price to its maximum, to the detriment of consumers, is avoidable through appropriate policy.[15] These issues are uppermost in understanding Smith's championship of laisser-faire. The mercantilist-state was observed to intervene on behalf of favored businesses, not broadly on behalf of the consumer public. (Natural monopoly affects only those few consumers who have both the "wealth and the fancy" of engaging into competition over the rarity at stake.) But the logic of Smith's policy prescriptions applies more broadly. Even if interventionist institutions like anti-trust are created to support consumers, they are likely to be "captured" by the narrower interest of business firms, and become a means of protecting incumbent firms, who are visible and influential, from new entrant competitors, who are invisible and without influence. Or perhaps the regulatory institution and the protected firms may naturally develop a common interest and capture each other.[16]

---

[15] Smith's thinking is illustrated by the proposition: you have to give in order to receive as in all voluntary exchange; interventions qualify this, allowing benefits to be received without giving.

[16] Competition is best understood as market, not production, rivalry. Competing firms may share single facilities as a joint venture. Thus, morning and evening newspapers may jointly own a printing press; several shippers might share docking facilities or flight runways.



*Exchange value.* The ratio at which commodities exchange for one another. In terms of a value standard, exchange value simplifies into price (nominal price being exchange value in terms of money).

*Monopoly price*. See natural price.

*Natural price*. Formally, "natural price" means minimum price (which emerges when supply is highly abundant relatively to demand), in contrast to "monopoly price", the maximum price (which emerges when supply is very scarce relatively to demand):

> "The price of monopoly is upon every occasion the highest which can be got. The natural price, or the price of free competition, on the contrary, is the lowest which can be taken, not upon every occasion indeed, but for any considerable time to­gether. The one is upon every occasion the highest which can be squeezed out of the buyers, or which, it is supposed, they will consent to give: The other is the lowest which the sellers can commonly afford to take, and at the same time continue their business." (A. Smith, 1776 [1904], Book I, Ch. VII, p. 63)

That Adam allows for price to be temporarily below the natural price is not in itself a contradiction to the meaning of natural price as the minimum price: in his characteristic realism, pushed at times to great detail, he allows for the producer to temporarily sell units at a loss (perishable goods like imported oranges), rather than incur the bigger loss of not selling at all. One may simplify the discussion by avoiding casual cases of sales at a loss, without sacrificing generality, since this exceptional case is easily included by a mere redefinition of the seller's reservation price (one for the current market period; another for long term sustainability). But Adam Smith also used natural price as a syno­nym for: (2) cost (see definition); (3) equilibrium price or even attractor price (hence his



oft-quoted gravitation metaphor); (4) "normal price" or "ordinary price". The justification for these other meanings is a general attitude among the classical economists, which Adam Smith inaugurated, and which consists of treating *free competition (entry)* as a norm, as if it is the normal state of affairs in all markets. Thus, Adam Smith explains price adjustment in disequilibrium assuming supply-demand imbalances around the natural price; also, his restriction of *effective demand* (see definition) to *"those who are willing to pay the natural price"*. This is not a definition of effectual demand, but merely a specification of it in the given context.

*Scarcity and water paradox*. The classical view on competition (and monopoly) is best articulated, not in terms of the number of buyers or sellers per se, but in terms of their ratio, which measures the scarcity of a commodity. The demand-and-supply ratio is ubiquitous in the classical literature, so much so that Ricardo noted: "The opinion that the price of commodities depends solely on the proportion of supply to demand, or demand to supply, has become almost an axiom in political economy." (Ricardo, 1817 [2004], ch. XXX, p. 382) Of course to think in terms of the ratio of supply and demand is essentially the same as to think in terms of their difference, which is more common today. Yet the absolute version of this ratio, which has no counterpart in modern theory, plays a pivotal role in the classical view of competition. Until the marginal revolution, indeed, scarcity is measured by the ratio between the overall number of potential units consumers need of a good and the total number of units of the good that potentially can be supplied; that is, in classical jargon, the ratio of absolute demand to absolute supply (as total potentials). Scarcity thus understood is an indicator of potential competition between suppliers and demanders in a market, and is therefore an indicator of the potential competitive price of a commodity: the scarcer a good (e.g. a diamond, where



absolute supply is small relative to absolute demand) the more competition will there be among the demanders of the commodity, hence the greater will be its equilibrium market price; on the other hand, the more abundant is a good, the more competition there is among the suppliers of the good, who, by underselling each other to supply the relatively fewer customers in need of the good (e.g. water which is abundant except in deserts), will bring the market price closer to its lowest possible value, the natural price (see definition).[17]

In place of this global, aggregate, objective measure of scarcity, the marginal school substituted a local, individual, subjective one: marginal utility, and its property of being a diminishing function of the quantity an individual possesses of a good.[18] The more units of a good an individual already has, the less valuable is an additional unit to the consumer. Later, after the ordinal turn in the neoclassical school, this subjective relation between value and scarcity (diminishing marginal utility) becomes irrelevant to ordinal value theory, and is replaced by a more elaborate one: diminishing marginal rates of substitution—the notion that the more units of a good an individual already has, the more units of this good he or she would be willing to exchange for another more desired commodity (Hicks, 1939 [1946] pp. 20-22).

*Use value*. The value a buyer is willing to pay for a commodity by virtue of the commodity's usefulness. As clarified by Jules Dupuit (1844, p. 343; 1849, p. 182), use-value is

---

[17] These classical conceptions connect readily to modern, and Hayekian notions of knowledge. Thus, the potential suppliers of anything are large or small depending on the extent of knowledge in the sense of knowledge-how. Diamonds are no longer a monopoly to the extent that today they can be manufactured, although demanders may perceive them as imperfect substitutes.

[18] Walras, who was familiar with the old definition of rareté through his father (Auguste Walras), adopted marginal utility as a substitute to it (Jaffé, 1972).



measured by maximum willingness to pay reservation price. J.S. Mill reached the same conclusion: "Value in use […] is the extreme limit of value in exchange", that is, price (1848 [1965], bk. 3, ch. 1, § 2, p. 457). Or: "the utility of a thing in the estimation of the purchaser, is the extreme limit of its exchange value." (1848 [1965], bk. 3, ch. 2, § 1, p. 462)

**7 A Mathematical Model for Chapter VII**

We turn finally to using Adam Smith's sketch of competitive price theory (clarified in Section 5) to state a mathematical model of the classical process of price formation based on a set of rivalrous market interactions that takes the forms of buyer-buyer out-bidding, seller-seller underselling, or buyer-seller higgling, where all buyers and sellers arrive in a market with maximum willingness to pay and minimum willingness to accept distribution function representations of individual demand and supply.[19]

This framework implements two methodological principles that seem to be at the center of a modern theory of price formation in the original spirit of the old school:

*Principle 1: Realism. Market behavior is founded on concepts that are observable and operational. Supply and demand are classically defined by an observable, operational, monetary value: the reservation price—the buyer's maximum willingness to pay and the seller's minimum willingness to accept.*

*Principle 2: Emergent rationality. Market interactions determine deep emergent properties that are the unintended consequences of people's actions, the results of human actions and not of human design.*

From these simple methodological principles, it should be possible to derive an alternative picture of a market economy that has an integrity distinct from the neoclassical theory. The second, equally fundamental, notion of market rationality as an emergent order (as the famous invisible hand metaphor conveys) is also lost in modern conventional price theory, in which the rationality of market outcomes rests, in final analysis, on the rationality of an idealized agent (Walrasian auctioneer, social planner, representative agent) and is not the consequence of privately informed traders whose interaction discovers prices that summarize the essence of their dispersed private information.

---

[19] Although the precise mathematical concept of a distribution was lacking in the classical era, a few authors of that era (notably Germain Garnier, J.-B. Say, and Jules Dupuit) came close to this formal representation of demand, which they identify with the distribution of buyers' wealth or willingness to pay, depicted as a pyramid (Garnier, 1796 [1846], pp. 195-196; Say, 1828 [1836], pp. 171-172, 175; Dupuit, 1844, p. 368).

Classical competitive price discovery in general, and Adam Smith's formulation in Chapter VII in particular, for a non-re-tradable good or service, can be derived from three basic assumptions that describe a simplified, reduced form, of our mathematical theory of price formation:

1. (Motivation) An individual is willing to trade, if there is any gain from trading (namely if there is surplus to be gained, given unit willingness to pay or demand values and willingness to accept or supply costs).
2. (The law of supply and demand) Price change and excess demand have the same sign.
3. (Short-side principle) Quantity traded is the minimum between quantity supplied and quantity demanded.

By virtue of the first assumption, the market supply and demand are represented fully as the cumulative distribution of sellers' unit costs and buyers' unit values, respectively.

Let the value and cost distributions be denoted respectively as:

$$D(v) = \#\{i : v_i \geq v\}. \tag{1}$$

$$S(c) = \#\{j : c_j \leq c\}. \tag{2}$$

Define the abundance of the good (the inverse of scarcity) or scarcity of the commodity by the ratio of the (maximum) number of units of the commodity available for supply, by the (maximum) number of units demanders need of the commodity:

$$\alpha = \frac{m}{n} = \frac{\max_x S(x)}{\max_x D(x)} = \frac{S(\infty)}{D(0)}. \tag{3}$$



Let $p(t)$ be the commodity's standing market price. The "large market" case is assumed merely because it allows us to illustrate the theory with familiar smooth-curve charts.[20]

The second assumption, the classical (dynamic) law of supply and demand, formally reads,

$$[D(p) - S(p)]\frac{dp}{dt} \geq 0. \tag{4}$$

Finally, the third assumption, the short-side principle, simply says that,

$$Q = \min(D, S). \tag{5}$$

Consider the following distance measure between the market price and the individual valuations of the good:

$$V(p) = \sum_{\{i: v_i \geq p\}} |v_i - p| + \sum_{\{j: c_j \leq p\}} |c_j - p|, \tag{6}$$

in which the notation means summation of all values $v_i \geq p$ and all costs $c_j \leq p$ the qualification being due to the fact that no units will be traded at a loss.

The overall surplus generated is obtained similarly, replacing the quantities supplied and demanded with the quantities traded.

It can be shown that the function $V$ is an integral of excess supply:

$$V(p) = V(0) + \int_0^p [S(x) - D(x)]dx. \tag{7}$$

In a large market, where this function is not only continuous but also smooth, we have, by the chain rule:

---

[20] A referee asks if it is not "dubious whether political economy before the marginal revolution such as Smith's can be theorized in differential calculus." However, none of our results depend on "large market" smoothness and differentiability. We use this limiting case to allow results to be expressed in familiar modern neoclassical terms. Only integration as summation is required, mathematically, which depends not on smoothness.



$$\frac{dV}{dt} = \frac{dV}{dp}\frac{dp}{dt} = [S(x) - D(x)]\frac{dp}{dt}. \tag{8}$$

Thus, by the law of supply and demand (4), this distance between price and the valuations is nonincreasing (technically, it is a Lyapunov function of competitive price dynamics):

$$\frac{dV}{dt} \le 0. \tag{9}$$

The property (9) has a fascinating interpretation that echoes Hayek's intuition about the informational function of a competitive market price.[21] It reveals the sum of information about consumers' needs, means, tastes, and producers' production capacities—dispersed information not in the reach of any single mind. Property (9) also means that the market price of a good evolves so as to reflect the traders' valuations and costs better and better, until the distance between the market price and the distribution of values and costs is minimized. A competitive equilibrium price is thus a generalized median of the traders' valuations.[22] Interestingly, this emergent informational optimization was first discovered, but neither appreciated nor generalized, in early experimental data as best explaining, empirically, the dynamics of laboratory markets and was referred to as the "minimum rent hypothesis" (V. L. Smith, 1962).

The three assumptions (1)-(3) imply the following result (illustrated in Figure 1 for a "large market"), which we state without proof.

---

[21] In the absence of a "large market", we have more generally: $\Delta V_t = V_{t+1} - V_t \le 0$, transaction by transaction, trader profits and welfare keep getting better and better, in the limit approaching the center of value where V is minimized.

[22] It is a generalized median in the sense of minimizing a generalized mean-absolute error function, namely the price-value distance function V.



***Theorem 1***: *Assume a competitive market (in the classical sense of underselling, outbidding, and higgling and bargaining, not price taking!). Assume no re-trading takes place (thus, no speculation). Then price converges to minimum price-value distance, maximum trade, and maximum surplus.*

Coming back to Ch. VII more specifically, the following result summarizes formally the key propositions of classical price theory intuitively derived so far (see also Figure 1).

***Theorem 2:*** *Consider a good traded in a large market. Then its competitive equilibrium price is an increasing function of its scarcity: that is, over some range, $p^* = f(\alpha)$ with $f'(\alpha) < 0.$ Moreover, the competitive price tends to a "natural value," if the good is extremely abundant, and to the monopoly value, if the good is extremely scarce: $p^* \to \min(c) \, as \, \alpha \to \infty \, and \, p^* \to \max(v) \, as \, \alpha \to 0.$ If all goods in the economy can be produced in abundant amounts, at proportional costs, using homogenous labor, then the natural general equilibrium[23] of the economy is a Leontief price system; hence all the goods would be priced according to the total labor involved in their production: $\mathbf{p^* = l(I - A)^{-1}},$ where $\mathbf{l}$ is the vector of direct labor requirement per unit of output and $\mathbf{A}$ is the matrix of direct material input requirements per unit of output (and $\mathbf{I}$ being the identity matrix).*

---

[23] That is, an equilibrium where all goods are traded at their minimum prices.



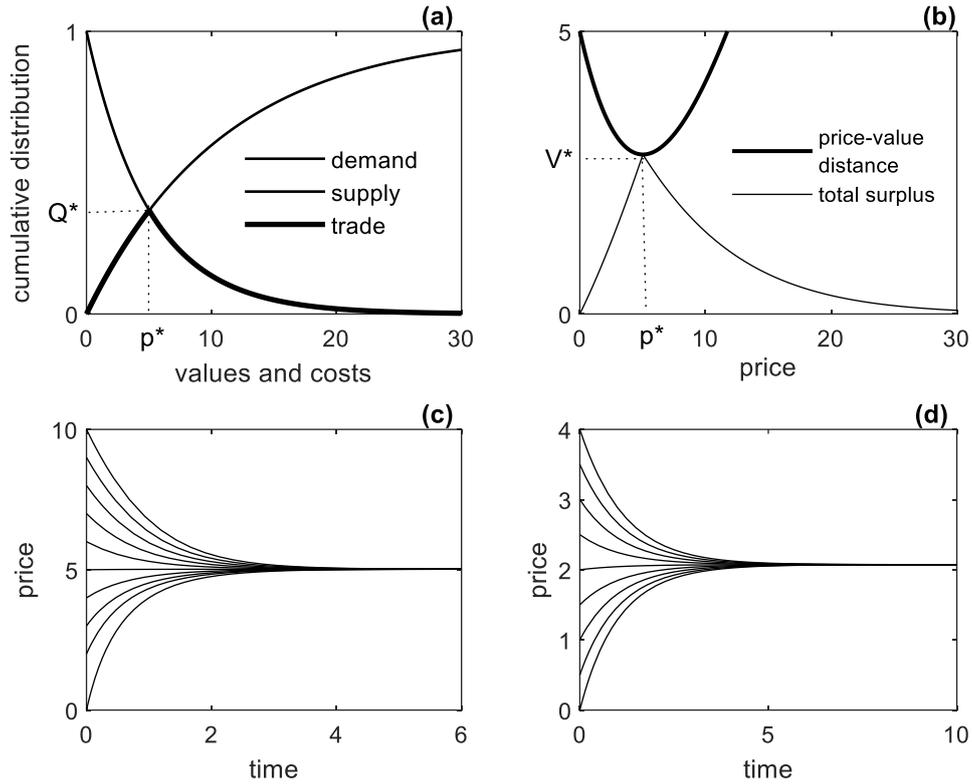

Figure 1: Price formation in a large market. The reservation prices are here expo-nentially distributed. In (a)-(c), mean(values)=5, mean(costs)=11, and scarcity ra-tio=1, making for equilibrium price p*≈ 5. Idem in (4), save for min(costs)=2 and scarcity ratio =0.01. In (c)-(d) are price trajectories for various initial conditions.

Proof. In a large-market, competitive equilibrium reduces to the traditional market-clearing concept, defined here by the equation $n\operatorname{prob}(v \geq p^*) = m\operatorname{prob}(c \leq p^*)$, or $\operatorname{prob}(v \geq p^*) = \alpha \operatorname{prob}(c \leq p^*)$. Since each probability being between 0 and 1, it follows that $0 \leq \operatorname{prob}(c \leq p^*) \leq 1/\alpha$ and $0 \leq \operatorname{prob}(v \geq p^*) \leq \alpha$. Thus $\operatorname{prob}(c \leq p^*) \to 0$ as $\alpha \to \infty$ and $\operatorname{prob}(c \leq p^*) \to 0$ as $\alpha \to 0$, implying, respectively, $p^* \to \min(c)$ and $p^* \to \max(v)$. Let $G(p) = \operatorname{prob}(v \geq p)$ and $F(p) = \operatorname{prob}(c \leq p)$. By the implicit function theorem, the equation $G(p^*) = \alpha F(p^*)$ implies that, over some interval in $[\min(c), \max(v)]$, $p^* = f(\alpha)$ with $f'(\alpha) = [G'(p) - \alpha F'(p)]/F'(p) < 0$. (Labor theory of value) The unit cost of commodity $k$ be written as $c_k = w_k l_k + \sum_h a_{hk} p_h$, where the



labor input is singled out: $l_k$ being the overall labor requirement and $w_k$ the average wage. Proportional costs means that both the matrix $\mathbf{A} = [a_{ij}]$ and the vector $\mathbf{l} = [l_i]$ are constant; homogenous labor implies a uniform wage rate $w_i = w$, which can be taken as the value standard, setting $w = 1$. All commodities are sold at their minimum willingness to accept means $p_i = c_i$, for $i = 1,...,n$, hence $p_i = \ell_i + \sum_k a_{ik} p_k$, which is a Leontief price system, whose existence and uniqueness are standard results of linear algebra [see, e.g., Meyer (2000, p. 681)].